\documentclass[10pt,aps,pra,twocolumn,a4paper,amsmath,amsthm,amssymb,showpacs,floatfix,superscriptaddress]{revtex4-1}
\usepackage{braket}
\usepackage{graphicx}
\usepackage{dcolumn}
\usepackage{dsfont}
\usepackage{epstopdf}
\usepackage{xcolor}
\usepackage{amsmath}
\usepackage[unicode=true]{hyperref}
\usepackage[normalem]{ulem}
\usepackage{siunitx}
\newcommand{\us}{\,\si{\micro\second}}

\begin{document}

\title{Optimized cross-resonance gate for coupled transmon systems}
\author{Susanna Kirchhoff}
\author{Torsten Ke\ss{}ler}
\author{Per J. Liebermann}
\email{perjliebermann@gmail.com}
\author{Elie Ass\'emat}
\author{Shai Machnes}
\affiliation{Theoretical Physics, Saarland University, 66123
	Saarbr{\"u}cken, Germany}
\author{Felix Motzoi}
\affiliation{Theoretical Physics, Saarland University, 66123
	Saarbr{\"u}cken, Germany}
 \affiliation{Department of Physics and Astronomy, Aarhus University,
DK-8000 Aarhus C, Denmark}
\author{Frank K. Wilhelm}
\affiliation{Theoretical Physics, Saarland University, 66123
	Saarbr{\"u}cken, Germany}
\date{\today}

\begin{abstract}
The cross-resonance (CR) gate is an entangling gate for fixed frequency superconducting qubits. While being simple and extensible, it is comparatively slow, at $160$\,ns and thus of limited fidelity due to on-going incoherent processes. Using two different optimal control algorithms, we estimate the quantum speed limit for a controlled-NOT (CNOT) gate in this system to be $10$\,ns, indicating a potential for great improvements. We show that the ability to approach this limit depends strongly on the choice of ansatz used to describe optimized control pulses and limitations placed on their complexity. Using a piecewise-constant ansatz, with a single carrier and bandwidth constraints, we identify an experimentally feasible $70$-ns pulse shape. Further, an ansatz based on the two dominant frequencies involved in the optimal control problem allows for an optimal solution more than twice as fast again, at under $30$\,ns, with smooth features and limited complexity. This is twice as fast as gate realizations using tunable-frequency, resonantly coupled qubits. Compared to current CR-gate implementations, we project our scheme will provide a sixfold speed-up and thus a sixfold reduction in fidelity loss due to incoherent effects.
\end{abstract}

\maketitle

\section{Introduction}
\label{sec:intro}
Circuit QED is a promising technology for quantum computing.  An important requirement for scalability of the architecture is high-accuracy implementation of two-qubit gates.  A leading candidate for resonator-mediated interaction between a pair of superconducting qubits is the so-called cross-resonance (CR) gate \cite{Paraoanu2006, Chow2011,Chow2012,Chow2014} which has been implemented experimentally with over $99$\% average gate fidelity \cite{Sheldon2016}.  The CR gate functions by driving one qubit at the resonant frequency of the other qubit, inducing dynamics in the latter across the connecting resonator, i.e.,~``cross-resonantly." The design avoids the complexity and noise sources that are present in low-frequency magnetic (flux) tuning of the qubit-qubit interaction \cite{Koch2007,groszkowski2011}. It also aims to improve on methods for high-speed addressing of specific two-qubit transitions, by utilizing the spatial addressability that comes from per-qubit control circuitry.

The primary impediments to high-fidelity operation currently come from two sources: incoherent errors, such as $T_1$ and $T_2$ processes, and coherent unitary errors, such as crosstalk \cite{Motzoi2013} and frequency crowding \cite{Theis2016}.  The main method to counteract the former is to shorten gate times as much as possible, but the increased spectral width can drastically increase unitary errors, especially from higher-order corrections to the perturbative model of the gate mechanism.

In this work we systematically optimize CR gate control pulses for best fidelity, by shortening gate times as much as possible to reduce incoherent errors, while avoiding the adverse effects of increased coherent errors. We employ a full Tavis-Cummings model \cite{tavis1968exact,tavis1969approximate} which eliminates many of the analytic simplifications that set bounds on analysis of the gate in the deeply non adiabatic regime.  Moreover, it is known that a careful analysis with regard to parametrization of the control sequence is required to efficiently tailor the control pulses to the constraints of the experimental apparatus \cite{Motzoi2011,GOAT-prl} and to ensure simple pulse shapes which permit experimental calibration of the pulse. By lifting constraints on pulse complexity, we numerically estimate the quantum speed limit (QSL) \cite{Levitin2009,Lapert2010,Caneva2009} for the gate.  However, the QSL is still dependent on other constraints imposed in the optimization problem and thus may also depend on the chosen parametrization of the control pulse. Therefore, we probe the relationship between the QSL and different physically meaningful parametrizations, which leads to a greater understanding of the limitations of the original cross-resonant control scheme and ultimately to improved control strategies.

This manuscript is organized as follows. Section \ref{sec:model} presents the theoretical model of the system. Section \ref{sec:method} introduces numerical methods. In Sec. \ref{sec:unconstrained-QSL} we show that the unconstrained quantum speed limit is much shorter than the currently prevailing strategy. Section \ref{sec:first-optim} describes the first optimization results and point outs the role of higher-frequency components. Section \ref{sec:QSL-Fourier} studies the QSL with the Fourier parametrization of the drive shape. Section \ref{sec:usable-result} explains the influence of the bandwidth constraint and proposes two highly practical solutions. Finally, in Sec. \ref{sec:conclusion} we make concluding remarks.

\section{System}
\label{sec:model}


Let us consider two transmon qubits coupled by a bus resonator. Each transmon is described as an anharmonic oscillator and the coupling to the resonator is described by an appropriately extended Jaynes-Cummings model \cite{Rigetti2012,Cross2015,Egger2014}. The qubits consist in the first two levels of these two anharmonic oscillators. As in Ref. \cite{Sheldon2016}, we assume that $\sigma_x$ and $\sigma_y$ type controls are available.
We further assume that the qubit frequencies can be calibrated by a quasistatic flux line. The aim of the latter is not to provide another time-dependent control function, but to statically shift the qubit frequencies to a value more favorable to the optimization process.  With the notations of \cite{Cross2015}, the Hamiltonian in the limit of low nonlinearity $\alpha$ \cite{khani2009} takes the form
\begin{align}
  H\left(t\right) =&  \sum\limits_{k=1}^2 \bigg(\omega_k b_k^\dagger b_k + \alpha_k b_k^\dagger b_k^\dagger b_k b_k  + g_k (a b_k^\dagger + a^\dagger b_k)  \nonumber\\
 &     +\sum_{l=1}^L \Omega_{k,l}(t)\cos\left(\omega^d_{k,l} t +\phi_{k,l} \right)  (b_k + b_k^\dagger)  \bigg)+ \omega_r a^\dagger a
\end{align}
where   $a^\dagger$, $b_1^\dagger$, and $b_2^\dagger$  are the creation operators for the cavity and the two transmons respectively; $g_k$ are the couplings between the resonator and the qubits; $\omega_r$, $\omega_1$, and $\omega_2$ are the respective frequencies of the cavity and the two transmons; $\Omega_{k,l}$ are the low-frequency envelopes of the microwave drive and $\omega^d_{k,l}$ are the carrier frequencies of the $L$ control drives with phase offset $\phi_{k,l}$. After moving to the rotating frame close to the frequency of the first qubit, and applying a rotating-wave approximation (RWA), the Hamiltonian takes the form
\begin{align}
  H_{\tiny\textsc{RWA}}\left(t\right) =
 &  \sum\limits_{k=1}^2 \left( \delta_k b_k^\dagger b_k + \alpha b_k^\dagger b_k^\dagger b_k b_k + g_k (a b_k^\dagger + a^\dagger b_k) \right. \notag \\
                & + \left.\Omega_k^x(t) (b_k + b_k^\dagger) + i \Omega_k^y(t) (b_k^\dagger - b_k) + f_k b_k^\dagger b_k \right)  \notag \\
               &+ \Delta a^\dagger a \, ,
\label{eq:main-ham}
\end{align}
where $\Delta$ is the detuning of the cavity from the principle (carrier) drive frequency of the controls, and $\delta_k$ are the detunings of the transmons. $f_k$ are introduced in the optimization as a mechanism for setting the transmon detunings and/or shifting of the principle drive frequency. The numerical simulations and optimizations presented below include the three first levels of the resonator and the three first levels of each transmon. System parameters are detailed in Table \ref{tab:main-ham-param}.

\begin{table}
\caption{Values of the parameters of the Hamiltonian in Eq. (\ref{eq:main-ham}), which has an RWA applied. These correspond to typical parameters for the dispersive regime of circuit QED.}
\label{tab:main-ham-param}
\begin{center}
\begin{tabular}{cc}
\hline\noalign{\smallskip}
Parameter & Value in GHz  \\
\noalign{\smallskip}\hline\noalign{\smallskip}
$\Delta /(2 \pi)$ & $0.4$  \\
$g_1 /(2 \pi)$ & $0.1$  \\
$g_2 /(2 \pi)$ & $0.1$  \\
$\alpha_{1,2} /(2 \pi)$ & $-0.32$  \\
$\delta_1 /(2 \pi)$ & $0.0$  \\
$\delta_2 /(2 \pi)$ & $-0.67$  \\
\noalign{\smallskip}\hline
\end{tabular}
\end{center}
\end{table}


\section{Numerical methods}
\label{sec:method}

The field of quantum optimal control (QOC) provides methodologies by which a quantum system may be driven to a desired state, or undergo a desired evolution, in a fast and efficient manner. With the emergence of quantum technologies \cite{Glaser2015} the significance and impact of these techniques has grown. Specifically, the requirement of very-high-fidelity gates for quantum computation and the complexity of the systems involved imply that approximate analytical solutions will not suffice, and numeric optimization theory must be applied to attain the desired process fidelities. Extensive research has gone into the problem of finding optimal driving of quantum systems. The field emerged in the mid-to-late 1980s with the first applications of QOC to chemical reactions and magnetic resonance imaging (MRI) \cite{TannorRice1985,optim6,Krotov2,Krotov1}, with experimental work continuing to this day \cite{optim-application1,optim-application2}. Since, the scope of QOC has widened considerably, with applications to attosecond physics \cite{atto1} and high harmonic generation \cite{optim-application3}, energy flow in biomolecules \cite{optim-application4},  and quantum computing \cite{optim-quan-comp1,optim-quan-comp2}, among others. QOC can be applied to both coherent and Markovian dynamics for state generation and other variants (see Refs. \cite{Machnes2011,Koch-Markovian}).

QOC methods can be roughly divided into two categories, gradient-free and gradient-based optimization, where the terms refer to the availability of the gradient of the goal measure to be minimized, with respect to the control parameters. With gradient-free methods, one samples the goal function at one or more points in the control-parameter space and deduces one or more new points for sampling, where the expectation is of an improved goal measure, and then repeats the process. This approach is simple and flexible and is the only possible procedure for closed-loop calibration. However, such methods converge very slowly compared to gradient-driven optimization, particularly when optimizing high-dimensional parameter spaces. The most well-known member of this class of optimizers is the Nelder-Mead algorithm  \cite{Nelder-Mead}, on which the quantum chopped random basis (CRAB) and dressed CRAB (dCRAB) methods \cite{CRAB,dCRAB} are based. Other gradient-free algorithms include the covariance matrix adaptation evolution strategy (CMA-ES) \cite{CMA-ES}, the simultaneous perturbation stochastic approximation (SPSA) \cite{spall1992multivariate}, and genetic algorithms, among others. And while approaches are often better at handling large parameter spaces and the presence of noise, they are still slow to converge compared to gradient-driven methods. 

When the gradient of the goal measure with respect to the control variables can be computed quickly (when compared to finite differences), gradient-based methods are preferred. These include the Krotov family of algorithms \cite{krotov1983iteration,Tannor1992,Koch-Krotov-Main} and the gradient ascent pulse engineering (GRAPE) \cite{GRAPE} method. Both are derived  from the variational formulation of the QOC task \cite{pontryagin1986mathematical}, where the Schr\"odinger equation is imposed via a Lagrange multiplier, which turns out to be a conjugate state evolving back in time. The method by which the control fields are updated in both the Krotov and GRAPE methods are defined using time-local expressions, implying a piecewise-constant (PWC) control ansatz, which may be detrimental in cases of bounded bandwidth. Combining Floquet theory with the variational approach also offers some advantages \cite{PRA-2013-Mintert}. The QOC method gradient optimization of analytic controls (GOAT) (see Ref. \cite{GOAT-prl} and detailed below) utilizes modified Schr\"odinger equations to compute the gradient and does not resort to variational calculous. We make use of GOAT below due to its simplicity and its flexibility of enforcing control constraints, such as bandwidth or power.
For a comprehensive review of QOC, see Refs. \cite{RabitzReview,Glaser2015}.

Given a system whose dynamics is described by the drift Hamiltonian $H_{0}$ and is subject to a set of control Hamiltonians $H_{k}$, the time-dependent Hamiltonian is
\begin{equation}
H\left(\bar{\alpha},t\right)=H_{0}+\sum_{k=1}^{C}c_{k}\left(\bar{\alpha},t\right)H_{k}\,,
\label{eq:Ham_tot}
\end{equation}
where $c_k$ are the control functions (ansatz), with details prescribed by the set of parameters $\bar{\alpha}$. One is free to chose any control parametrization, e.g., the Fourier basis,
\begin{equation}
c_{k}\left(\bar{\alpha},t\right)=\sum_{j=1}^{m}A_{k,j}\cos\left(\omega_{k,j}t+\phi_{k,j}\right),
\label{eq:Fourier-controls}
\end{equation}
with
\begin{equation}
\bar{\alpha}=\left\{ A_{k,j},\omega_{k,j},\phi_{k,j}\right\} _{k=1\ldots C,j=1\ldots m}.
\label{eq:p-bar}
\end{equation}
In the systems investigated in this work we have found the Fourier and Erf parametrizations to produce pulse shapes with low parameter counts. The goal function to minimize is defined as the infidelity (projective $SU$ distance) between the desired gate,  $U_{\textrm{goal}}$, and the implemented gate, $U\left(T\right)$,
\begin{equation}
g\left(\bar{\alpha}\right):=1-\frac{1}{\textrm{dim}\left(U\right)}\left|\textrm{Tr}\left(U_{\textrm{goal}}^{\dagger}U\left(T\right)\right)\right|\,,\label{eq:goal}
\end{equation}
where $U\left(t\right)$ is the time-ordered ($\mathbb{T}$) evolution operator
\begin{equation}
U\left(\bar{\alpha},T\right)=\mathbb{T}\exp\left(\int_{0}^{T}-\frac{i}{\hbar}H\left(\bar{\alpha},t\right)dt\right).\label{eq:Upt}
\end{equation}

QOC methods can be roughly divided into two categories: gradient-free and gradient-driven methods. The latter require the gradient of the goal function with respect to the control parameters, $\partial_{\bar{\alpha}}g\left(\bar{\alpha}\right)$, and are much faster provided this gradient can be computed efficiently. Gradient-free methods are appropriate for closed-loop calibration and when the gradient cannot be determined with ease. Gradient-driven QOC methods require a gradient-driven search method over the parameter space, $\bar{\alpha}$. Both GRAPE and GOAT (discussed below) utilize a standard algorithm for that purpose, limited-memory Broyden-Fletcher-Goldfarb-Shanno (L-BFGS) \cite{L-BFGS}.

\textit{GRAPE.} When $H\left(\bar{\alpha},t\right)$ is taken to be a PWC function, the method of choice for QOC is GRAPE \cite{GRAPE,Machnes2011}. We note that when PWC is used as an approximation of a smooth control field, it introduces non-negligible inaccuracies, which are discussed in Ref. \cite{GOAT-prl}. Let us enumerate the time slices $j\in\left\{1,\ldots,M\right\}$, each of duration $\Delta t_j$, $U_{j}=\exp\left(-\tfrac{i}{\hbar}\Delta t_j H_{j}\right)U_{j-1}$, where $H_j:=H_0 + \sum_{k=1}^{C}c_{j,k} H_k$ and $U_0:=\mathcal{I}$, $U\left(T\right):=U_M$. Here $\bar{\alpha}=\left\{c_{j,k}\right\}$. The gradient of the goal function [Eq. (\ref{eq:goal})], $\partial_{\bar{\alpha}}g\left(\bar{\alpha}\right)$, can be computed using the chain rule and $\partial_{\bar{\alpha}}U\left(\bar{\alpha}\right)$. Noting the $j,k$ component of $\bar{\alpha}$ appears only in $U_j$, the gradient of $U\left(T\right)$ is computed by $\partial_{\bar{\alpha}_{j,k}}U\left(T\right)$ $=$ $\left(\Pi_{b=M}^{j+1}U_b\right)\left(\partial_{\bar{\alpha}_{j,k}}U_j\right)\left(\Pi_{a=j-1}^{1}U_a\right)$. The expression to compute $\partial_{\bar{\alpha}_{j,k}}U_j$ $=$ $\partial_{c_{j,k}}exp\left(-\tfrac{i}{\hbar}\Delta t_j H_{j}\right)$ $=$  $\partial_{c_{j,k}}exp\left(-\tfrac{i}{\hbar}\Delta t_j \left(H_0 + \sum_{k=1}^{C}c_{j,k} H_k\right)\right)$ is rather cumbersome, and requires a full eigendecomposition of $H_j$ (see Ref. \cite{Machnes2011} for details). While computationally expensive, the eigendecomposition can be leveraged to perform exponentiation, propagation, and propagator gradients with little additional numerical effort. Therefore, GRAPE satisfies one of the conditions for a good gradient-based QOC method---the gradient of the propagator can be computed efficiently. The gradient of the propagator is then used to compute the gradient of the goal function, feeding into the L-BFGS search algorithm, which seeks to minimize the goal function over the $\bar{\alpha}$ parameter space.

\textit{GOAT.}
Consider the gradient of the goal function (\ref{eq:goal}) with respect to $\bar{\alpha}$,
\begin{equation}\label{eq:goal-func-grad}
\partial_{\bar{\alpha}}g\left(\bar{\alpha}\right) =-\textrm{Re}\left(\frac{g^{*}}{\left|g\right|}\frac{1}{\textrm{dim}\left(U\right)}\textrm{Tr}\left(U_{\textrm{goal}}^{\dagger}\partial_{\bar{\alpha}}U\left(\bar{\alpha},T\right)\right)\right)\,.\\
\end{equation}
Generally, $U\left(\bar{\alpha},T\right)$ does not have a closed form [see Eq. (\ref{eq:Upt})], and therefore  $\partial_{\bar{\alpha}}U\left(\bar{\alpha},T\right)$ cannot be computed directly. As $U$ evolves by the equation of motion (EOM) $\partial_t U\left(\bar{\alpha},t\right) = -\frac{i}{\hbar}H\left(\bar{\alpha},t\right) U\left(\bar{\alpha},t\right)$, we may take the derivative of the $U$ EOM with respect to $\bar{\alpha}$, swapping derivation order, resulting in a system of coupled EOMs:
\begin{equation}
\partial_{t}\left(\begin{array}{c}
U\\
\partial_{\bar{\alpha}}U
\end{array}\right)=-\frac{i}{\hbar}\left(\begin{array}{cc}
H & 0\\
\partial_{\bar{\alpha}}H & H
\end{array}\right)\left(\begin{array}{c}
U\\
\partial_{\bar{\alpha}}U
\end{array}\right).\label{eq:joint-eom}
\end{equation}
$\partial_{\bar{\alpha}}H$ is computed using the chain rule and Eqs. (\ref{eq:Ham_tot}) and (\ref{eq:Fourier-controls}). The coupled time evolution of the propagator (a single equation of motion) and its gradients ($C\times m$ equations---as per the dimension of $\vec{\alpha}$) may be performed by any mechanism for ordinary differential equation (ODE) integration that is accurate and efficient for time-dependent Hamiltonians, such as adaptive Runge-Kutta. A gradient-driven search over the parameter space is performed using L-BFGS.


\begin{figure}
\includegraphics[width=\columnwidth]{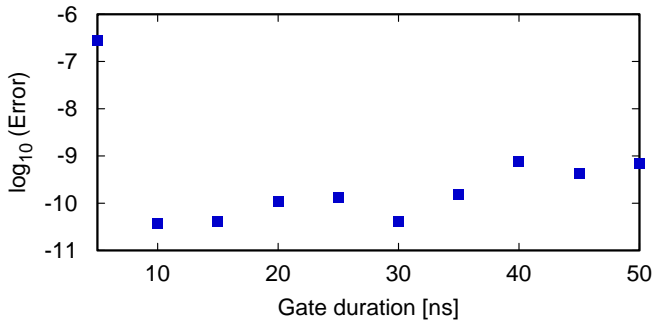}
\caption{We observe a $5$- to $10$-ns quantum speed limit (QSL) for the system described by Eq.~(\ref{eq:main-ham}) and Table~\ref{tab:main-ham-param}, using arbitrary amplitude piecewise-constant controls with a high sampling rate ($\leq 0.1$ns) and no filter. The unitary error (infidelity; $1-\Phi_\mathrm{goal}$) of the optimized controlled-NOT (CNOT) gates is plotted as a function of gate duration, with each point representing the average error for multiple optimizations starting at random initial pulses. See Sec. \ref{sec:unconstrained-QSL} for a complete discussion.}
\label{fig:QSLunfilt}
\end{figure}

\section{Unconstrained Quantum Speed Limit}
\label{sec:unconstrained-QSL}

We first look for the QSL for the system described by Eq.~(\ref{eq:main-ham}) and Table~\ref{tab:main-ham-param}, by using the least-constrained parametrization, i.~e.~, PWC, with no amplitude bounds and high resolution---$500$ time slices for each gate duration $t_\mathrm{g}$. Assuming the parametrization is sufficiently flexible, the QSL observed results from the physics of the system. One standard method to probe the QSL numerically is to plot a measure of the gate fidelity as a function of the gate duration \cite{Egger2014,Soerensen2016}. For different gate durations we perform multiple GRAPE optimizations, starting each with a different random initial guess pulse. The average final gate errors, defined as $g=1 - \Phi_{\textrm{goal}}$, are shown in Fig. \ref{fig:QSLunfilt}. We observe a clear jump between 5- and 10-ns gate times, which indicates the presence of a QSL. Interestingly, the speed limit is more than an order of magnitude shorter than the gate time presented in Ref. \cite{Chow2011}, suggesting a potential for reduction of incoherent errors by an order of magnitude. As no constraints are imposed at this stage, these optimal control shapes are beyond the capabilities of most existing experimental setups. However, this is likely to change in the near future, with adoption of AWGs with bandwidths above $10$GHz, such as those demonstrated in the circuit QED setup of Ref. \cite{Raferty2017}, and FPGA control logic. Even without such hardware, by taking into account the constraints of the experiment implementation allows us to implement gates in times which, while higher than the unconstrained QSL, are significantly shorter than the original cross-resonance gate, as discussed in the following sections.

\section{Smooth control in the time domain}
\label{sec:first-optim}

To reduce the complexity of the generated pulse sequence, allowing effective calibration, we optimize control pulses in the PWC time domain, smoothed by a Gaussian filter. Based on the analytical methods presented in Ref. \cite{Paraoanu2006}, and further refined and experimentally verified in Ref. \cite{Chow2011}, a cross-resonance gate is generated in a two-transmon system by applying driving on one qubit at the frequency which is resonant with the second qubit. Let us consider the effective coupling $J_{\textrm{eff}}$ between the two qubits that quantifies the effective interaction mediated by the resonator. In the case where $J_{\textrm{eff}}$  is small compared to the detuning between the qubits, $\delta_2$, a drive at frequency $\tilde{\omega_2} = \omega_2 - (J_{\textrm{eff}})^2/\delta_2$ generates dynamics that can be described by an effective Hamiltonian of the form

\begin{equation}
H_{\textrm{eff}} = u_1^{\textrm{eff}} \sigma_1^{z} \otimes \sigma_2^{x} + u_2^{\textrm{eff}}  \sigma_1^{x}\otimes \mathds{1} \, ,
\label{eq:cross-res-ham}
\end{equation}
where $\sigma_i$ are the Pauli operators and $u_i^{\textrm{eff}}$ denote the relative scaling of the effective interaction. The $ZX$ interaction present in this effective Hamiltonian can generate a CNOT gate directly, up to local rotations~\cite{Rigetti2010}.

The performance of the initial guess is further improved by limiting the effect of the second term in Eq.~(\ref{eq:cross-res-ham}).   Here we assume this spectral constraint (i.e., suppressing the spectral weight of the second term) can be satisfied by using a second off-phase quadrature $\Omega^y$ of the control envelope set proportional to the derivative of $\Omega^x$ and inversely proportional to the qubit frequency separation $\delta_2$, as in the derivative removal by adiabatic gate (DRAG) method \cite{Motzoi2009,Motzoi2013}.

These analytical techniques would not be sufficient to obtain high-fidelity gates at very short times for the model at hand due to the complexity of the level structure. However, the initial guess is found to be relevant enough to be located in the basin of attraction of a higher-fidelity solution, which allows the GRAPE algorithm to converge on a solution with suitably smooth features.

\begin{figure}[htb]
\includegraphics[width=\columnwidth]{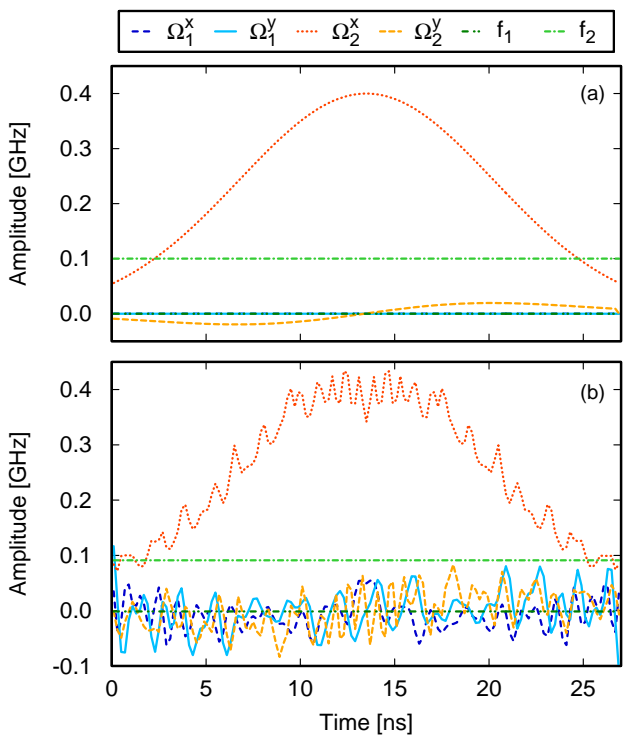}
\caption{High-fidelity CR control pulse sequence found with a sampling rate of $5$Gs/s. (a) Initial control guess pulse conforming to the analytical ansatz (\ref{eq:initial-guess-1}). (b) Optimized controls giving $\Phi_{\textrm{goal}} = 0.9999$. The pulses are relatively smooth and short (27 ns) but contain a large degree of complexity. See Sec. \ref{sec:first-optim} for full details.}
\label{fig:first-optim-bounded}
\end{figure}

Thus, we choose the following initial control functions:
\begin{align}
\Omega^x_1(t) & = 0 \quad  \Omega^y_1(t)  = 0 \notag\\
\Omega^x_2(t) & = 0.4 \exp\left(-\frac{(t - \mu)^2}{2 \sigma^2}\right) \quad \mu=\frac{t_g}{2}, \, \sigma = \frac{t_g}{4}  \notag\\
\Omega^y_2(t) & = \frac{1}{\delta_2} \dot{\Omega}^x_1(t) \, , \quad f_1  = 0 \, , \quad f_2  = 0.1 \, ,\notag \\
\label{eq:initial-guess-1}
\end{align}

where the amplitudes are given in gigahertz, and $f_1$ and $f_2$ are constant, but tunable, frequency offsets relative to the drive frequency of qubits 1 and 2, respectively. Control amplitudes are bounded, and the sampling rate is lowered to $0.2$ ns. When the optimization is run with the parameters in Table \ref{tab:main-ham-param}, and a slightly increased gate time of $t_g = 27$ ns is set, we achieve a gate infidelity of $10^{-4}$, assuming no incoherent processes. The pre- and postoptimization drive shapes are shown in Fig. \ref{fig:first-optim-bounded}. We observe that the strong cross-resonant drive remains after the optimization, which indicates that the CR scheme is still the main physical mechanism in play.  The pulse induces complicated dynamics to obtain high-fidelity in a very short time, suggesting it may be possible to reduce gate times further, achieving an order of magnitude acceleration compared to Ref. \cite{Sheldon2016}, thus reducing the decoherence by a similar fraction and outperforming gates with tunable qubit frequency architectures \cite{Kelly2015}.  The pulse is simplified compared to those used to identify the gate QSL, and the sampling rate is well within the current technological capacities of next generation microwave generators \cite{Raferty2017}. In Sec. \ref{sec:QSL-Fourier} we further simplify the parametrization of the pulse, which allows for experimental implementation with a minimal amount of overhead.

\section{Quantum Speed Limit with the Fourier ansatz}
\label{sec:QSL-Fourier}

It is extremely beneficial to arrive at quantum control sequences that are parameterized by a small number of parameters.  This simplicity can greatly aid in analytical, numerical, and experimental reproducibility of the solution, by allowing easy closed-up in situ calibration (or tune-up) of the control sequences. To gauge how the complexity of the pulses increases as we get close to the QSL, we use a parametrization based on the number of Fourier components of the pulse, though any other set of parameters more appropriate for the given control task could be substituted. Each control is parametrized by a truncated Fourier series that is fed  to a product of two sigmoids to enforce the global bound and a smooth start and finish. The amplitude coefficients, frequencies, and phases are optimized using the GOAT algorithm \cite{GOAT-prl}, a recent gradient base algorithm that is capable of handling any analytic ansatz and can easily accommodate multiple and varied constraints.

\begin{figure}[htbp]
\includegraphics[width=\columnwidth]{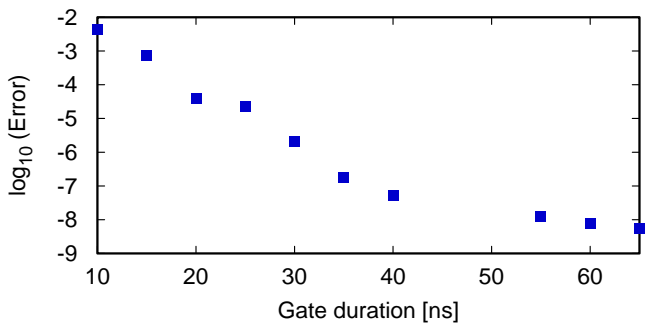}
\caption{Determination of the QSL using 167 Fourier components per control. The Errors ($1-\Phi_\mathrm{goal}$) averaged over optimized pulses found for different random initial conditions are plotted for different gate durations.  Errors are larger and decay more slowly when compared to a time-domain parametrization containing roughly 800 (PWC) control points, depicted in Fig. \ref{fig:QSLunfilt}.}
\label{fig:QSL2}
\end{figure}

We begin with a large number of Fourier components and probe the QSL. The number of components is chosen to offer roughly the same number of parameters to describe the drive shape as we had in the PWC case of Fig. \ref{fig:QSLunfilt}. We observe in Fig. \ref{fig:QSL2} that the QSL is less sharp than what was observed with the PWC description, which illustrates the clear influence of the choice of the control representation on the control landscape. Then, we iteratively remove the Fourier component with the smallest amplitude and reoptimize. The process terminates when simpler controls are unable to reach the minimum gradient threshold. In this case, component count was lowered down to only nine components. However, this reduction is at the cost of an increased gate time of $70$\,ns and the appearance of some very high frequencies.  This offers a hint to explain why the unconstrained piecewise optimization manages to converge to a gate error of $10^{-10}$, whereas the spectral optimization with a smaller frequency range converges only to $10^{-3}$. It seems that high-frequency components are necessary for the fine-tuning needed to achieve a high accuracy. Moreover, this could also be a sign that the Fourier ansatz is not the most efficient for this system, and one may wish to try a few other analytic ansatzes. The GOAT package would be well suited for such study.

Nonetheless, limiting the number of Fourier components to 167 we find that the minimum time is around 15\,ns to obtain $10^{-3}$ infidelity,  which is consistent with the PWC case and thus indicates this is likely the QSL for such a regime of control parameters, regardless of the chosen control parametrization. The dependence of QSL on the number of control parameters (15\,ns for 167 components vs 70\,ns for 9 components) demonstrates that the quantum speed limit is not only a function of time but also a function of pulse complexity. Thus, the Fourier basis is a good choice to minimize pulse complexity, as might be suited to in situ tuning of pulse parameters. This representation is also natural for enforcing bandwidth constraints. Lastly, it lends itself to the generation of highly simplified pulses by an iterative process of reducing the number of Fourier components. We do have to keep in mind, however, that microwave-pulse-shaping technology is typically in the time domain.

\section{Bandwidth-constrained pulses}
\label{sec:usable-result}

The PWC parameterization, with high-resolution controls, provides a theoretic lower bound to gate times. In practice, time resolved optimizations are limited by the AWG's bandwidth, as well as other filtering induced by system components. We therefore optimize the pulses for an AWG with a finite time resolution of $1$\,ns, fine steps of 0.2\,ns, buffers of 4\,ns in duration at the beginning and the end, and filtering of the signal, consistent with earlier work \cite{Motzoi2011}. Filtering is applied via a Gaussian window function with the standard deviation $\sigma=0.4$\,ns, i.e., a bandwidth of 331\,MHz. We consider two control ansatzes: one in which the controls utilize a single carrier frequency (the standard CR scheme), and one in which two carrier frequencies are employed.

\begin{figure}[htbp]
\includegraphics[width=\columnwidth]{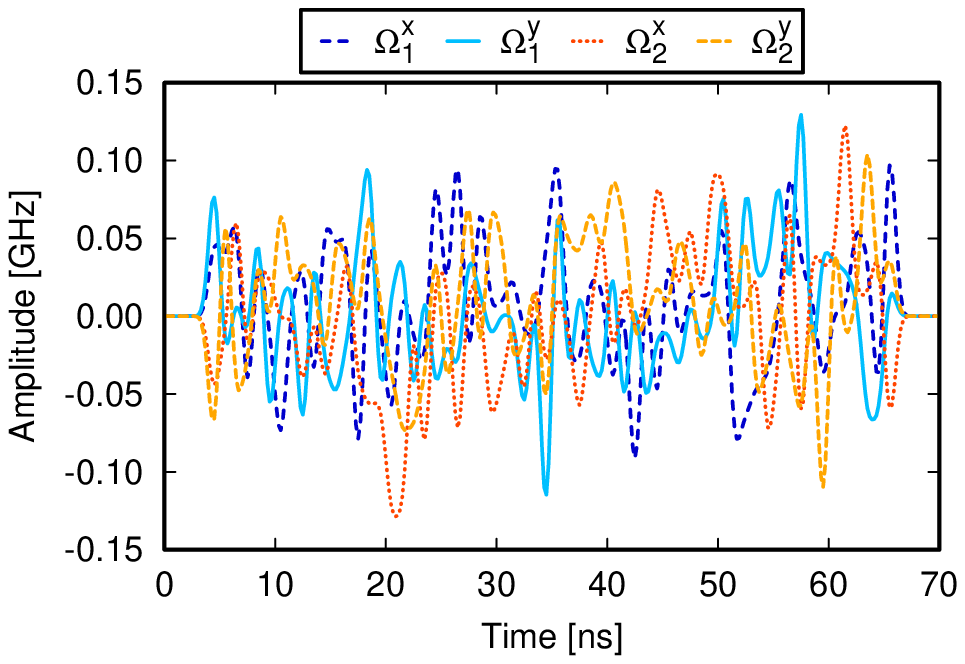} \\
\includegraphics[width=\columnwidth]{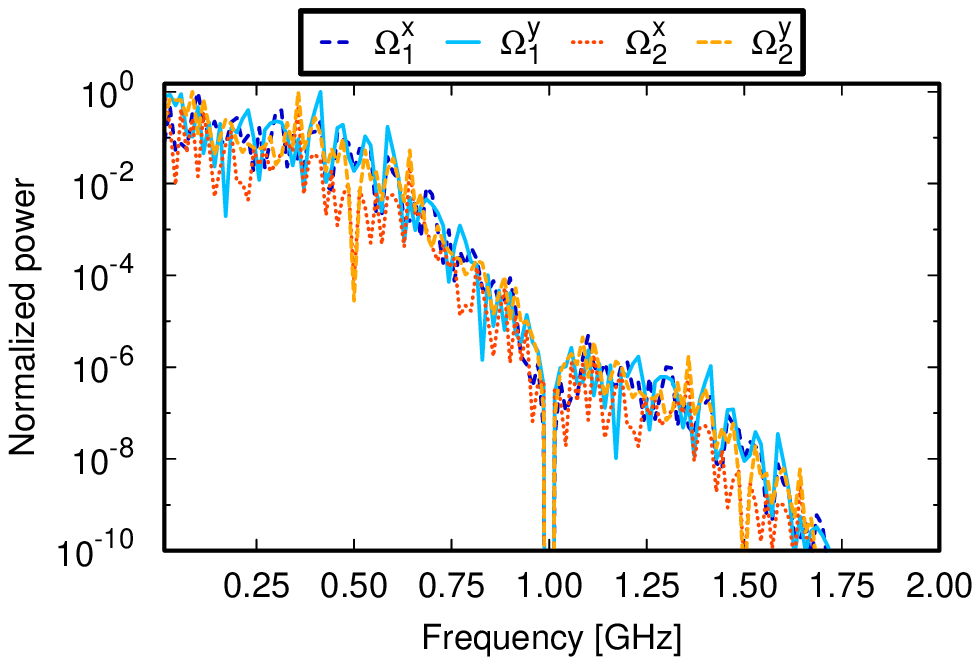}
\caption{Pulse shapes and spectra of the optimal drive shape for filtered PWC ansatz with a gate time of 70\,ns and a single carrier frequency. A fidelity of 0.999 is achieved using a 331-MHz-bandwidth Gaussian filter. }
\label{fig:Filtered-controls}
\end{figure}

When a single carrier frequency is used, a minimum gate time of $70$\,ns is achieved, with a fidelity of 0.999.  GRAPE optimization generated the controls and spectra shown in Fig. \ref{fig:Filtered-controls}.  Clearly visible is the reduced bandwidth, effected by the control filtering. As a result, the pulse requires significantly more time than the QSL. However, it is still less than half the time required by current implementations of the CR gate \cite{Sheldon2016}, implying only half as much fidelity will be lost to incoherent processes.

Moving to two carrier frequencies and retaining the bandwidth limitations on the AWGs, we see a drastic reduction in the gate time. The intuition to this scheme stems from Sec. \ref{sec:first-optim}, where one can identify two principle components in the frequency spectrum of the controls appearing in Fig. \ref{fig:first-optim-bounded}.  As can be expected, the two principle frequencies are proximate to the resonance frequencies of the two qubits. More precisely, the qubit frequency at $0.67$ GHz is shifted by $f_2$ to $0.57$ GHz, which is then bifurcated by the Rabi splitting $2g$ to $0.47$ and $0.67$ GHz. Our control pulses take the following forms:

\begin{figure}[htb]
\includegraphics[width=\columnwidth]{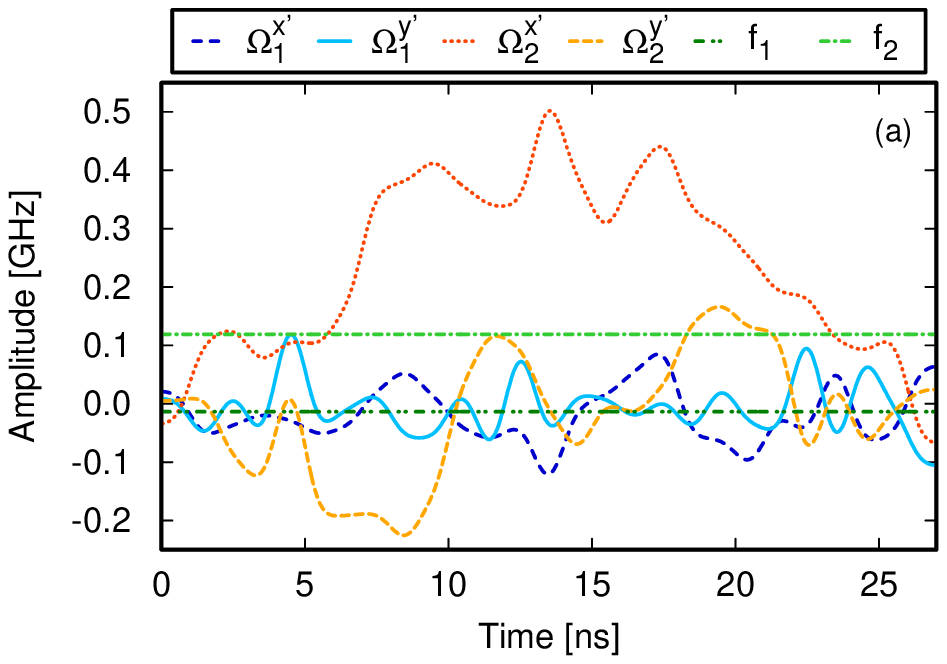} \\
\includegraphics[width=\columnwidth]{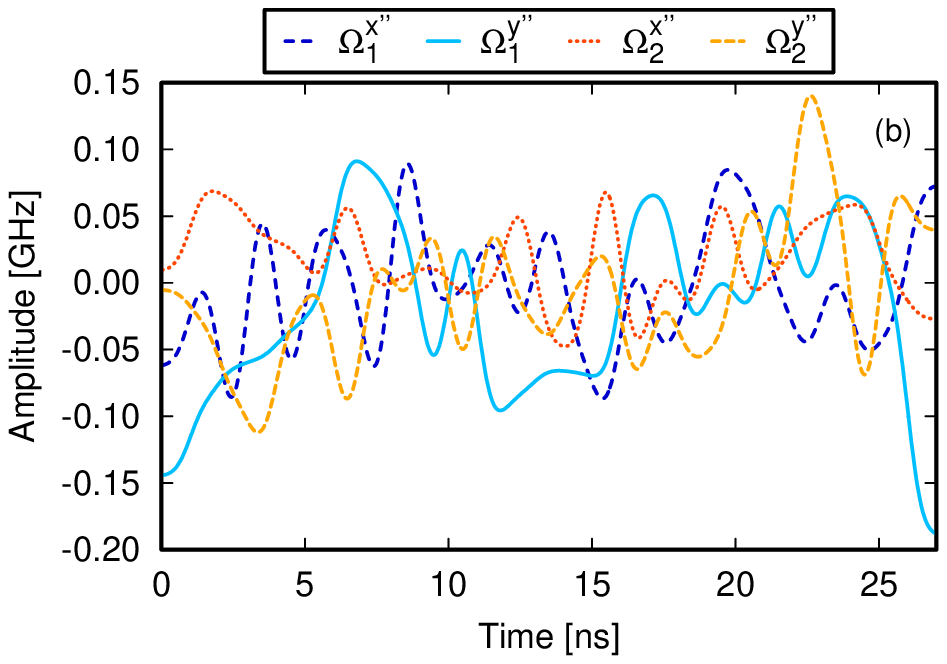}
\caption{Optimization results for control pulses using two carrier frequencies and a bandwidth filter, achieving $0.9999$ fidelity when incoherent processes are ignored. (a) Controls at qubit 1 frequency. (b) Controls at qubit 2 frequency ($\delta\pm g$).}
\label{fig:twotone}
\end{figure}

\begin{align}
\Omega^x_1(t) & = \Omega^{x'}_1(t) + \cos((\delta +g)t)\Omega^{x''}_1(t) \label{eq:initial-guess-2} \notag \\
 \Omega^y_1(t) & = \Omega^{y'}_1(t) + \cos((\delta +g)t)\Omega^{y''}_1(t) \notag \\
\Omega^x_2(t) & = \Omega^{x'}_2(t) + \cos((\delta -g)t)\Omega^{x''}_2(t) \notag \\
 \Omega^y_2(t) & = \Omega^{y'}_2(t) + \cos((\delta -g)t)\Omega^{y''}_2(t)
\end{align}
where the quotes symbols ($\,''$) denote the new (AWG) control functions, and $\delta = \delta_2 + f_2$ is the qubit separation. This doubles the number of functions to optimize. A Gaussian filter with a bandwidth of $331$\,MHz is added to account for the distortion of the PWC control functions by wave-form generators \cite{Motzoi2011}. The smooth controls, in addition to being a constraint of the system, improve optimization convergence speed and may help with experimental imperfections and unforeseen low-pass filters in the system. The pulse complexity is reduced not only by decreasing the sampling rate and the bandwidth of the pulses but also by the very short gate time of 27\,ns, which indicates that the total number of control points needed is reasonably small, on the order of $100$.

The optimization is carried out with coarse pixels of $1$\,ns and a fine time step of $0.05$ ns and reaches a final fidelity of $0.9999$. The control functions optimized $\Omega^{i'}_j$, $\Omega^{i''}_j$, and $f_j$ are shown in Fig. \ref{fig:twotone}. The constant value of the frequency detuning is also optimized but its value appears to be stuck in a local minimum and does not evolve significantly during the optimizations, suggesting additional improvement may be achieved by fine-tuning the choice of drive frequencies near the qubit transitions.

\begin{figure}[htb]
  \includegraphics[width=\columnwidth]{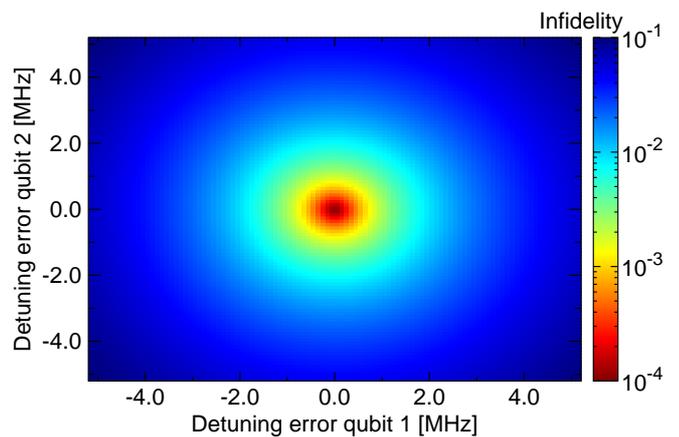}
\caption{Infidelity of the CR gate as a function of the miscalibration of qubit 1 and qubit 2 frequencies, 
with other model parameters specified by Table \ref{tab:main-ham-param}. The control pulses are as depicted in Fig. \ref{fig:twotone}.}
\label{fig:miscalib}
\end{figure}

In Fig. \ref{fig:miscalib}, fidelity dependence on $f_1\pm\epsilon_1$ and $f_2\pm\epsilon_2$, is plotted. We notice qubit-driven detuning errors $\epsilon_i$ on the order of up to about 2\,MHz lead to a drop of fidelity by 2 orders of magnitude. Experimentally, miscalibrations are expected to be much smaller than this amount.

\begin{figure}[htb]
  \includegraphics[width=\columnwidth]{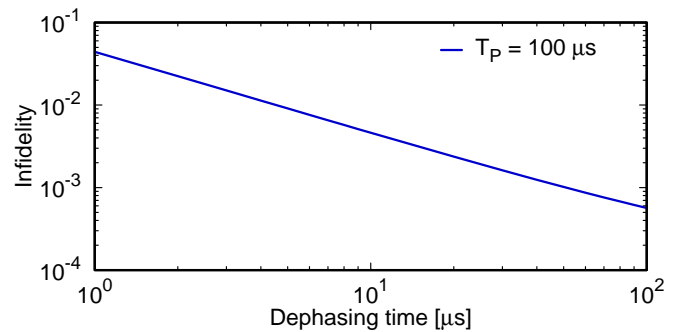}
\caption{Infidelity of the CR gate as a function of equal dephasing and decoherence rates. Model parameters follow Table \ref{tab:main-ham-param}, and control pulses follow Fig. \ref{fig:twotone}.}
\label{fig:dephasing}
\end{figure}

In Fig. \ref{fig:dephasing}, we examine the fidelity's dependence on dephasing time, with dephasing and relaxation $T_1=T_2$, and the cavity decay rate
set to $T_P=100\us$. The full evolution is then given by the master equation
\begin{equation}
   \dot{\rho}=-i [H,\rho] + \gamma_j\sum_j A_j \rho A_j^\dagger - \frac{1}{2} A_j^\dagger A_j \rho - \frac{1}{2} \rho A_j^\dagger A_j\,.
\end{equation}
The Lindblad operators are $b_i$ for relaxation, $\sqrt{b_i^\dagger b_i}$ for pure dephasing, and $a$ for the cavity decay. We see that for typical values of experimental dissipation losses ($>100\,\mu$s), the error is limited only by the precision of our unitary 
optimization and not by additional non-unitary losses, validating our estimates for the CR quantum speed limit.

\begin{figure}[t]
\includegraphics[width=\columnwidth]{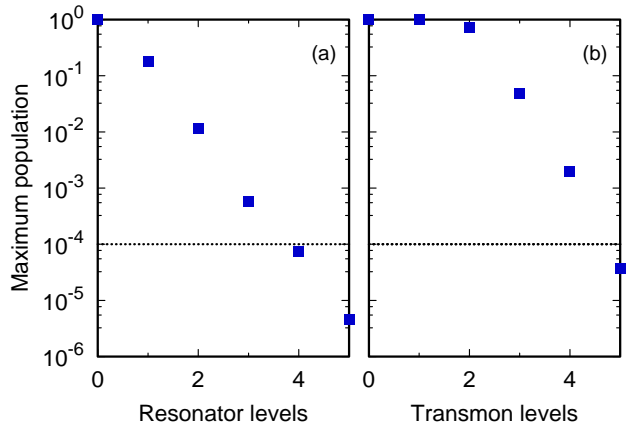}
\caption{Maximum transient population leakage to the higher levels of the resonator (a) and the transmons (b) during the dynamics driven by the optimal controls described in Fig. \ref{fig:twotone} and the model parameters detailed in Table \ref{tab:main-ham-param}. Note the logarithmic scaling.}
\label{fig:higher-levels}
\end{figure}

Finally, we test the validity of the truncation of the Hilbert space. The optimized control functions were applied to a larger Hilbert space utilizing six levels for the resonator and each transmon. 
The maximum population reached during the dynamics is noted in Fig.~\ref{fig:higher-levels}. For the transmons, we take the maximum over both qubits
$M_r = \max_t(p_r(t))$ and $M_{\textrm{Transmons}} = \max_{t;i\in 1,2}(p_{r_i}(t))$. 
The population of the resonator's third level, as well as the qubits' fifth level, remain below $10^{-3}$. Unsurprisingly, the transmon is leakier than the resonator by just about an order of magnitude: Leakage to level three is approximately $10^{-3}$ for the resonator and $10^{-2}$ for the qubits.

\section{Discussion}
\label{sec:conclusion}

In this work we have demonstrated the very significant benefits which can be derived from applying quantum optimal control to the problem of quantum gate generation and more specifically to superconducting gates. We have chosen to focus on CR gates as one of the leading architectures for superconducting gates, which benefit from not requiring the overhead or noise sources present in tunable-frequency qubits.

The current state of the art for the CR CNOT gate has an infidelity of $0.009\pm±0.002$ \cite{Sheldon2016}. One of the primary reasons for 
the imperfect fidelity is the $160$-ns duration of the gate, allowing incoherent processes to induce an infidelity of $0.004$ ($T_1 \approxeq 40\us$ and $T_2 \approxeq 5\us$). Therefore, to improve fidelity significantly, one must shorten the  gates (in parallel with efforts to increase coherence times and reduce coherent errors).

We note that the model used for the CR circuit is not perfect, as with any experiment: some aspects of the model are unknown (e.g., random defects in the substrate coupling to the circuit), and some characterization gaps and parameter drifts 
are unavoidable. Therefore, every proposed pulse sequence will have to be calibrated in a closed-loop tune-up process---a gradient-free search over the space of pulse parameters with the goal of minimizing infidelity. Such calibrations are 
practical only when the pulses are described by a small number of parameters. Therefore, practical pulses must have a simple description.

It is critical to determine the limits of achievable performance with any given circuit design, CR gates included, as this affects the decision of whether new circuits need to be developed to achieve fidelity goals. We have therefore employed several quantum optimal control techniques (GRAPE and GOAT) to determine the quantum speed limit and to design simple bandwidth-constrained control pulses which implement significantly faster CR gates.

Specifically, we have shown that the speed limit of the system is between $5$ and $10$\,ns, implying an incoherence limit of below $5\times10^{-4}$. Unfortunately, the pulse shapes required to reach such speeds are too complex to reliably implement in the experiment and far too complex to calibrate.

We therefore explored alternate routes: the first follows the standard control scheme, where only the control qubit is driven, and the second requires contemporaneous driving with two carrier frequencies. Both approaches yielded significant improvement over the state of the art.

With a single carrier, we have achieved a CNOT gate in only $70\,$ns. A further dramatic acceleration can be achieved when a second carrier is introduced: We have identified a pulse sequence which implements a CNOT gate in only $27\,$ns. For such short durations, incoherent effects induce less than $10^{-3}$ infidelity. We note that the control fields used are low-pass-filtered to the current circuit's control bandwidth and are therefore directly implementable. Moreover, both pulses are described by less than $100$ parameters and are therefore calibratable. To reduce parameter count further, one may employ the GOAT optimal control method, to allow additional control ansatzes to be explored.

Our exploration of the quantum speed limit further shows that, unlike the case where there are no direct qubit drives and the natural (perfect) entangler to use is the iSWAP \cite{goerz2015optimizing,Goerz2017}, a CNOT gate is instead a good fit for architectures where such drive lines do in fact exist.  Moreover, in comparison to the exhaustive parameter search for a global QSL without a direct line, \cite{Goerz2017}, in our limited local search with a direct line we are able to cut in half the global QSL with our bandwidth-constrained pulses. Nonetheless, one may still be able to use the insights from the former to find an even faster operating regime for the transmons in the latter case, notably by moving the transmon frequencies towards the quasi-dispersive straddling qutrits (QuaDiSQ) regime~\cite{Goerz2017}.

This work also motivates further use of this already prominent gate, being in fact even faster than the entangling gates used in the 
tunable-frequency implementations, which suffer from extra noises originating at the additional flux-tuning circuitry.

To conclude, we have shown that the application of quantum optimal control to the cross-resonance superconducting CNOT gate can reduce 
pulse duration from $160$ to $27$\,ns, using control sequences which are well within experimental capabilities, are 
described by a small number of parameters, and are therefore calibratable. Thus, we demonstrate the potential of reducing incoherent effects fivefold, significantly improving gate fidelity.

\begin{acknowledgments}
E.A. acknowledges support from the Alexander von Humboldt Foundation. S.K. and S.M. acknowledges funding from the IARPA through 
LogiQ Grant No. W911NF-16-1-0114. P.J.L. and F.K.W. acknowledge support from the European Union through the ScaleQIT project.
\end{acknowledgments}

\bibliography{literature}

\end{document}